\documentclass[aps,twocolumn,pre,showpacs,10pt,floatfix]{revtex4-1}
\usepackage{graphicx}
\usepackage{amsmath}
\usepackage{amsfonts}
\usepackage{mathtools}
\usepackage{physics}
\usepackage{subfigure}
\usepackage{xcolor}

\begin{document}
\newcommand{\nwc}{\newcommand}
\nwc{\vs}{\vspace}
\nwc{\hs}{\hspace}
\nwc{\la}{\langle}
\nwc{\ra}{\rangle}
\nwc{\lw}{\linewidth}

\nwc{\nn}{\nonumber}
\nwc{\pd}[2]{\frac{\partial #1}{\partial #2}}
\newcommand{\be}{\begin{equation}}
\newcommand{\ee}{\end{equation}}
\newcommand{\bea}{\begin{eqnarray}}
\newcommand{\eea}{\end{eqnarray}}
\newcommand{\PP}{\mathcal{P}}

\title{Inference of entropy production for periodically driven systems}

\author{Pedro E. Harunari$^1$, Carlos E. Fiore$^2$, and Andre C. Barato$^3$}
\affiliation{$^1$Complex Systems and Statistical Mechanics, Department of Physics and Materials Science, University of Luxembourg, L-1511 Luxembourg City, Luxembourg\\
$^2$Instituto de F\'isica da Universidade de S\~ao Paulo, 05314-970 S\~ao Paulo, Brazil\\
$^3$Department of Physics, University of Houston, Houston, Texas 77204, USA
}

\parskip 1mm
\def\d{{\rm d}}
\def\Ps{{P_{\scriptscriptstyle \hspace{-0.3mm} s}}}
\def\MF{{\mbox{\tiny \rm \hspace{-0.3mm} MF}}}
\def\ts{T_{\textrm{sig}}}
\def\tos{T_{\textrm{osc}}}
\begin{abstract}
The problem of estimating entropy production from incomplete information in stochastic thermodynamics is essential for theory and experiments. 
Whereas a considerable amount of work has been done on this topic, arguably,  most of it is restricted to the case of nonequilibrium steady states driven by a fixed 
thermodynamic force. 
Based on a recent method that has been proposed for nonequilibrium steady states, we obtain an estimate of the entropy production based on the statistics
of visible transitions and their waiting times for the case of
periodically driven systems.
The time-dependence of transition rates in periodically driven systems produces several differences in relation to steady states, which is reflected in the entropy production estimation.
More specifically, we propose an 
estimate that does depend on the time between transitions but is independent of the specific time of the first transition, thus it does not require tracking the protocol. Formally, this elimination of  the time-dependence
of the first transition leads to an extra term in the inequality that involves the rate of entropy production and its estimate.   
We analyze a simple model of a molecular pump to understand the relation between the performance of the method 
and physical quantities such as energies, energy barriers, and thermodynamic affinity. Our results with this model indicate that the emergence of net motion in the form of a probability current in the space of states is a necessary condition for a relevant estimate of the rate of entropy production.

\end{abstract}

\maketitle

\section{Introduction}

Stochastic thermodynamics \cite{jarz11, seif12, broe15} is a modern theoretical framework that generalizes thermodynamics to systems that can be small, i.e., 
made only of a few degrees of freedom, and out of equilibrium. The success of this theory is also due to the fact that such small 
systems have become accessible in experiments. Examples include, single molecules such as molecular motors, colloids 
and quantum dots \cite{cili17}. A main observable of interest in stochastic thermodynamics is the rate of entropy production, which quantifies 
the thermodynamic cost of an out of equilibrium system.    

It is often the case in an experiment that not all states of the mesoscopic system are accessible.  Therefore, the inference of properties of the system, 
such as its rate of entropy production, from partial information is a fundamental theoretical problem in stochastic thermodynamics. In particular, while well-controlled experiments with full information about the system have provided beautiful connections between modern theory and experiment \cite{coll05,blic09,beru12,kosk14,diet15,kuma20}, connecting stochastic thermodynamics with less controlled experiments in molecular biophysics remains a major challenge, which is closely related to the problem 
of inferring properties of the system from partial information.

Much work on the problem of inferring the entropy production and related observables has been done. The thermodynamic 
uncertainty relation \cite{bara15a,ging16,piet16} has been used to infer entropy production and related observables from the fluctuations of 
a current \cite{bara15,piet16b,piet17b,li19,vu20,dech21}. The rate of entropy production can also be estimated from the Kullback-Leibler divergence between 
the probability of a coarse-grained trajectory and the probability of the respective reversed trajectory  \cite{rold10,rold12,muy13,mart19}. Another approach to estimate the
 entropy production is to assume that the visible dynamics corresponds to a hidden Markov process and use the observable data to infer the underlying Markov process
\cite{ehri21,skin21,skin21a}. More generally, the related topic of coarse-graining in stochastic thermodynamics has been widely studied \cite{raha07,pugl10,espo12,mehl12,bo14,knoc15,shir15,bisk17,seif20,teza20}. Most of these works on estimation of entropy production are suitable for steady states,
however, some works have recently addressed time-dependent cases \cite{degu24,sing23,lee23, otsu22, koyu20}. 

Particularly important for this work is a method to estimate the rate of the entropy production from the statistics of the sequences and 
times between a few visible transitions proposed independently in two recent papers  \cite{meer22,haru22}.  This approach is promising as it 
is reasonable to expect that the statistics of the time between transitions, i.e., the intertransition time, is accessible in an experiment.
These two references concentrate on systems in nonequilibrium steady states, which are driven by constant thermodynamic forces.

In contrast to nonequilibrium steady states, periodically driven systems are driven by an external periodic protocol. 
Examples of periodically driven systems include cyclic heat engines \cite{schm08,blic12,bran15,mart17,kris16,datt22} and
 artificial molecular machines \cite{raha08,cher08,maes10,raha11,mand14,verl14,asba15,astu08,erba15}. 
It is known that periodically driven systems and nonequilibrium steady states can display substantial differences \cite{raz16,bara16}, 
therefore,  the inference method for steady states from \cite{meer22,haru22} prompt the following questions. 
How is the method extended to periodically driven systems? What are the differences in the application of the
method between steady states and periodically driven systems?

In this paper we provide answers to these questions. We show that  
 the rate of entropy production can be inferred from the statistics of sequences and intertransition times between a few visible transitions in periodically driven systems.
 We find several important differences in relation to steady states. Since transition rates are time-dependent, the statistics of transition pairs depend on the intertransition time and the
 initial time of the first transition.  We overcome this issue by developing an estimator that only depends on the intertransition time. The formal elimination of this dependence on the time of the first transition leads to an inequality between the estimator and the real rate of entropy production that contains an extra term. 
 We also find an estimator that depends on both the intertransition time and the time of the first transition. Even though this bound provides a 
 better estimate of the rate of entropy production, compared to the one that only depends on intertransition times, in practice, it requires one 
 histogram of intertransition times per each considered initial time, which can be computationally expensive.  

The estimator of the entropy production involves two different distributions of the intertransition time, one associated with the original forward protocol and the other 
 associated with the time-reversed protocol. In contrast to steady states that only involve one distribution.  Hence,  the application of the method requires  two ``experiments''
 one with forward protocol and another with the time-reversed protocol.
 
A simple three-state model with time-dependent energies and energy barriers \cite{ray17} is used as a proof 
of concept. We obtain the following results with this model. First, for a unicyclic network of states, while a similar method gives 
the exact value of the entropy production for steady states \cite{meer22,haru22}, 
our estimate only provides a lower bound on the rate of entropy production for periodically driven systems. 
Second, in contrast to steady states, periodically driven systems can have entropy
production in the absence of a net current in the space of states.  For this case of absence of net motion,  
the estimator for entropy production gives a numerical result compatible with zero, which suggests that net motion 
is an essential feature for our estimator to provide information about the rate of entropy production.

An analytical procedure to calculate the distribution of the probability density of the intertransition time for steady states 
was introduced in \cite{meer22,haru22, sekimotoFPT}. This procedure maps the problem of calculating the intertransition time distribution into the problem 
of calculating the time-dependent probability of an absorbing state of a certain auxiliary process. 
We generalize this procedure to periodically driven system, providing a pathway to obtain our estimator analytically.

The paper is organized as follows. In Sec. \ref{sec2}, we introduce the key quantities analyzed here and define
the three-state model. Our numerical results showing the application of the method to
the specific model are shown in Sec. \ref{sec3}. The inequality that involves the rate of entropy production, its 
estimator and an extra term is proved in Sec. \ref{sec4}.  
The analytical method to determine the intertransition time distribution is discussed in Sec. \ref{sec5}. 
We conclude in Sec. \ref{sec6}.

\section{Observable transitions, intertransition time statistics and inference of entropy production}
\label{sec2}
\subsection{Markov processes with time-periodic transition rates}
\label{sec21}
Here we consider Markov processes with a finite number of discrete states denoted by $i$ and $j$. The time-dependent transition rate from state $i$ to state
$j$ is denoted $w_{ij}(t)$. This transition rate is time-periodic with period $T$, i.e.,  $w_{ij}(t)=w_{ij}(t+T)$. The probability to be in state $i$ at time $t$ 
is written as $P_i(t)$ and follows the master equation
\begin{equation}
\frac{d}{dt}P_i(t)=\sum_{j\neq i} \left[P_j(t)w_{ji}(t)-P_i(t)w_{ij}(t)\right].
\label{EqMaster}
\end{equation}

We are interested in the long time limit $t\to\infty$, for which the distribution $P_i(t)$ becomes time-periodic with period $T$. This long time limit solution 
of the master equation is denoted by $P_i(t)$ for the remainder of this paper. In stochastic thermodynamics we restrict to rates with the property that if 
$w_{ij}(t)\neq 0$ then $w_{ji}(t)\neq 0$. 

The rate of entropy production is given by \cite{seif12}
\begin{equation}
\sigma\equiv \frac{1}{T}\int_0^T dt\sum_{i,j} P_i(t)w_{ij}(t)\ln\frac{w_{ij}(t)}{w_{ji}(t)}.
\label{eqentr}
\end{equation}
This physical quantity can be expressed as the sum of terms that are a product of thermodynamic flux and the respective thermodynamic affinity with the 
use of the generalize detailed balance relation \cite{bran15,ray17}. We discuss this expression for the entropy production for the particular model 
we introduce in the following section.

\subsection{Model definition}
\label{sec22}
\begin{figure}
\includegraphics[width=\columnwidth, trim = {0.4cm 0 0.1cm 0}, clip]{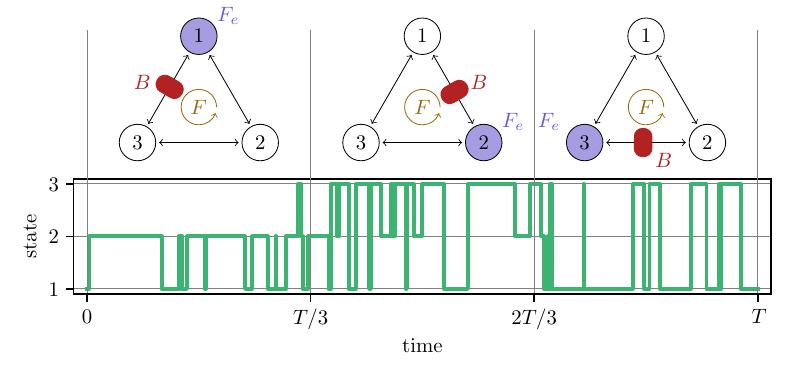}
\vspace{-2mm}
\caption{Upper panel: periodic protocol for the three-state model. The energy $F_e$ and the energy barrier $B$ rotate clockwise in each third part of the period.  Lower panel: example of a stochastic trajectory with transitions between states and waiting times.}   
\label{fig1} 
\end{figure}

As a proof of concept we consider the following model for a molecular pump depicted in Fig. \ref{fig1}. 
A similar model, with a stochastic protocol instead of the deterministic protocol 
considered here has been analyzed in \cite{ray17}. The model has three states, which we label as $i=1,2,3$.
There is one energy $F_e$ and one energy barrier $B$, while the remaining energies and energy barriers are zero. 
The periodic protocol is piece-wise, with the period divided into three parts. For the first part, corresponding to $0<t<T/3$, 
the energy of state $1$ is $E_1(t)=F_e$ and the energy barrier between states 3 and 1 is $B_{31}(t)=B$. The other two energies and two energy barriers are 
$0$. For the second  part, corresponding to $T/3<t<2T/3$,  $E_2(t)=F_e$,  $B_{12}(t)=B$, and the other energies and energy barriers are zero. 
For the third  part, corresponding to $2T/3<t<T$,  $E_3(t)=F_e$,  $B_{23}(t)=B$, and the other energies and energy barriers are zero.

We here set Boltzmann's constant and temperature to $k_B=T=1$, where $T$ represents temperature and not period only in this equation, throughout.  
The transition rates for this model are given by 
\begin{equation}
w_{ii+1}(t)= k\textrm{e}^{E_i(t)-B_{ii+1}(t)}
\end{equation}
and
\begin{equation}
w_{i+1i}(t)= k\textrm{e}^{E_{i+1}(t)-B_{ii+1}(t)}
\end{equation}
Due to periodic boundary conditions, for $i=3$, we set $i+1=1$. The parameter $k$ sets the speed of the rates. 

We now discuss the sources of entropy production in this model.  Work is exerted  on 
the system due to the time-dependence of the energies. The entropy production is equal to the average work and can be written 
as 
\begin{equation}
\sigma= F_eJ_e,
\end{equation}
\begin{equation}
J_e= \frac{1}{T}\sum_{i=1}^3\left[P_{i+1}(iT/3)-P_{i}(iT/3)\right].
\end{equation}
This expression for $\sigma$  is equivalent to Eq. \eqref{eqentr}  \cite{ray17}. It can be understood as follows. Let us consider the term $i=1$ 
in the summation, corresponding to the time $T/3$ of the protocol. At this time if the system is at state $1$, with probability  $P_1(T/3)$, 
it will lose an energy $F_e$ since the energy of the system changes from $F_e$ to $0$. 
If the system is at state $2$, with probability  $P_2(T/3)$, it will gain an energy $F_e$. The same reasoning applies to $i=2$ and $i=3$.

The energy barrier $B$ does not appear in the expression for $\sigma$ explicitly, it appears only implicitly as it affects the probabilities $P_i(t)$. 
Even if $B=0$ the rate of entropy production can be non-zero. However, an energy barrier $B$ is a necessary condition to create net motion
in the clockwise direction, as explained in the literature on  ``no-pumping theorems'' \cite{raha08,cher08,maes10}. This net motion is quantified by the average current
\begin{equation}
J=\frac{1}{T}\int_0^T dt\sum_{i=1}^{3}\frac{J_{ii+1}(t)}{3}.
\label{eqcurr}
\end{equation}
In other words, when $B = 0$, $J$ also vanishes. However, even with $J = 0$, the periodically driven system remains out of equilibrium when $Fe \neq 0$, resulting in a positive entropy production. This behavior contrasts sharply with steady state systems, where nonzero entropy production is always associated with currents. As shown below, the presence of this net motion quantified by $J$ in periodically driven systems is essential for our estimate of the rate the entropy production. If $J=0$
 our estimate is much smaller than the entropy production in general, providing little to no information.

We also consider the case of the presence of a fixed thermodynamic affinity $F$ that would drive the system to a non-equilibrium steady state in the absence
of a periodic protocol. The transition rates are modified to 
\begin{equation}
w_{ii+1}(t)= k\textrm{e}^{E_i(t)-B_{ii+1}(t)-F/3}
\label{EqRates1}
\end{equation}
and
\begin{equation}
w_{i+1i}(t)= k\textrm{e}^{E_{i+1}(t)-B_{ii+1}(t)},
\label{EqRates2}
\end{equation}
the negative sign in $-F/3$ in the first equation implies that a positive $F$ leads to a force in the counter-clockwise direction.

In the presence of this fixed affinity $F$ the entropy production in Eq. \eqref{eqentr} becomes
\begin{equation}
\sigma= F_eJ_e-FJ,
\label{EqEnt1}
\end{equation}
where $J$ is the current in Eq. \eqref{eqcurr}. This general expression for the entropy production in terms of currents $J_e$ and $J$ and affinities $F$ 
and $F_e$ has been obtained in \cite{ray17}. The minus sign in the term $FJ$ comes from the fact that $F$ points in the counter-clockwise direction and $J$ points in the clockwise direction.  

This model can also operate as an engine when the term $F_eJ_e$ is positive and the term $-FJ$ is negative. 
In this regime $F_e$ together with the energy barrier $B$ creates a clockwise current that does work against the counterclockwise force $F$ \cite{raha11,ray17}.  
We also analyze the estimator of the entropy production in this regime.

\subsection{Intertransition time} 
\label{sec23}

An example of a stochastic trajectory is shown in Fig. \ref{fig1}. This trajectory is a sequence of transitions and waiting times between transitions. We denote a jump, or transition, from state $i$ to state $j$ as $\ell=ij$. The reversed jump from $j$ to $i$ is denoted by $\bar{\ell}$.  The set of visible transitions is denoted by $\mathcal{L}$. For instance,
for the three-state model we have a total of six possible transitions, two between each pair of states. If we only have access to the two transitions between the pair of states $1$ and $2$, then two of the six transitions are part of the set $\mathcal{L}$.

\begin{figure}
\includegraphics[width=0.8\columnwidth]{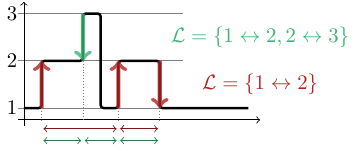}
\vspace{-2mm}
\caption{Sequence of intertransition time for $\mathcal{L}= \{1\to2,2\to1\}$ and for $\mathcal{L}= \{1\to2,2\to1,2\to3,3\to2\}$ . For this particular trajectory, 
there are two intertransition times for the first case and three intertransition times for the second case.}  
\label{fig2} 
\end{figure}

The distribution $\PP_{\ell',\tau+t_0\vert \ell,t_0}$ is the probability density that a transition $\ell'$ occurs at time $\tau+t_0$ given that transition $\ell$ occurred at the initial time  $t_0$ and that no other visible transition in $\mathcal{L}$ occurs before $\tau+t_0$, where $\tau$ is the intertransition time.
This dependence on the time of the first transition $t_0$ is a consequence of the time-dependent transition rates. An important fact about $\PP_{\ell',\tau+t_0\vert \ell,t_0}$ is its dependence on the set of visible transitions $\mathcal{L}$, which is illustrated in Fig. \ref{fig2} for a particular example. 
In this figure there are three transitions and two intertransition times if only the transitions between state $1$ and $2$ are visible and there are four transitions and three intertransition times if the transitions between state $2$ and $3$ are also part of $\mathcal{L}$. 
To simplify notation we do not write this dependence explicitly in $\PP_{\ell',\tau+t_0\vert \ell,t_0}$.

Our objective here is to infer the rate of entropy production from sequences and intertransition time statistics of a few visible transitions. 
Calculating the histogram associated with $\PP_{\ell',\tau+t_0\vert \ell,t_0}$ from a trajectory can be a hard task since it
depends on the intertransition time $\tau$ and the initial time $t_0$. A more practical method would be to only keep track
of the intertransition times in the trajectory, without keeping track of the time for the first transition. In practice, if we only keep track of the intertransition times,
we will average out over all possible initial times $t_0$. This average is over the probability 
$\PP_{t_0|\ell}$, which is the conditional probability density of sampling time $t_0\in [0,T]$ given that transition $\ell$ occurred. 
Based on this reasoning  we consider the quantity 
\begin{equation}\label{eq:psi}
\psi_{\ell'|\ell}(\tau)\equiv T^{-1}\int_0^T dt_0 \PP_{\ell',\tau+t_0\vert \ell,t_0} \PP_{t_0\vert \ell}.
\end{equation}
In other words, the quantity $\psi_{\ell'|\ell}(\tau)$ is particularly convenient from a practical perspective. 
If we simply count the number of transitions within a certain time-interval for the pair $\ell'|\ell$ in a trajectory, without accounting for the 
initial time of the first transition, we obtain the histogram associated with $\psi_{\ell'|\ell}(\tau)$.

The conditional probability density $ \PP_{t_0\vert \ell}$ can be written in terms of the transition rates $w_{ij}(t_0)$ and the long time limit 
probability $P_i(t_0)$ that is the solution of the master equation. If we denote $\ell=ij$ then 
\begin{equation}
 \PP_{t_0\vert \ell}= \frac{P_i(t_0)w_{ij}(t_0)}{\int_0^T dt_0 P_i(t_0)w_{ij}(t_0)}.
 \label{EqPtgivenl}
\end{equation}
In Sec. \ref{sec5} we provide a method to calculate $\PP_{\ell',\tau+t_0\vert \ell,t_0}$ and, consequently,  $\psi_{\ell'|\ell}(\tau)$ analytically.

\subsection{Reversed protocol} 
\label{sec24}

In order to estimate the entropy production from a few visible transitions in periodically driven systems we also need the  
statistics of the intertransition time associated with the reversed protocol. The mathematical definition for the reversed 
protocol is given by the following equation for the transition rates 
\begin{equation}
w^{\dagger}_{ij}(t_0)\equiv w_{ij}(T-t_0).
\end{equation}
Physically, for the model in Fig. \ref{fig1} the reversed 
protocol corresponds to the sequence of pictures showing the position of the energy and energy barrier in reverse order.

In order to calculate the probability density for the intertransition time associated with the reversed protocol $\psi^{\dagger}_{\ell'|\ell}(\tau)$,
we also need to generate a trajectory with the reversed protocol. From a practical perspective, in order to estimate the rate of entropy production with our method there is a need to generate trajectories from two 
experiments, one with the forward protocol and another with the backward protocol.

\subsection{Inference of entropy production from the statistics of the intertransition time}
\label{sec25}

Our estimator for  average rate of entropy production $\sigma$ is given by 
\begin{equation}
\label{vEPR_FPT}
\hat{\sigma} =K \sum_{\ell, \ell' \in \mathcal{L}} \int_0^\infty \mathrm{d}t \psi_{\ell' \vert \ell}(\tau)  \PP_\ell \ln \frac{\psi_{\ell' \vert \ell} (\tau)}{\psi^\dagger_{\overline{\ell} \vert \overline{\ell}'}(\tau)},
\end{equation}
where the activity $K$ is the average number of visible transitions in $\mathcal{L}$ per time and $P_\ell$ is the probability that an observable transition is $\ell$, irrespective of the time of its occurrence. These quantities $K$ and  $\PP_\ell$ can be written in terms of
the long-time solution of the master equation $P_i(t_0)$, they are given by,
\begin{equation}
K= T^{-1}\int_{0}^{T}dt_0\sum_{ij\in\mathcal{L}}P_i(t_0)w_{ij}(t_0)
\end{equation}
and
\begin{equation}
\PP_\ell= K^{-1} T^{-1}\int_{0}^{T}dt_0P_i(t_0)w_{ij}(t_0),
\end{equation}
where $\ell=ij$. We can obtain this estimate from a trajectory (and a second trajectory with the reversed protocol) where we can only observe a few visible transitions and the 
intertransition times. The activity $K$ is evaluated by simply counting the total number of transitions and dividing by the total time, $\PP_\ell$ can be obtained by counting the number of transitions $\ell$ and dividing by the total number of transitions,  $\psi_{\ell' \vert \ell}(\tau)$ can be obtained from a histogram of the intertransition times from the trajectory, and, finally, $\psi^\dagger_{\overline{\ell} \vert \overline{\ell}'}(\tau)$ can be similarly obtained with the only difference that the reverse protocol is applied.

\section{Results for case study}
\label{sec3}

In this section, we  show our numerical study of entropy production and the estimator for the model defined in Sec.~\ref{sec22}. We observed the estimate $\hat{\sigma}$ 
below the real rate of entropy production $\sigma$ in all our numerical results within this model. However, the formal inequality connecting 
both quantities contains an extra term $X$ and reads $\sigma\ge \hat{\sigma}-X$, where the extra term fulfills $X\ge 0$. This inequality is proved in the next section. 
In Fig. \ref{fig3} and Fig. \ref{fig4} we also illustrate it by plotting $\hat{\sigma}-X$, which can be negative.

The numerical results shown in the figures in this Section were obtained via Monte Carlo simulation with discretized time, where the time-step, which is much shorter than the characteristic time of the fastest jump, is small enough such that no relevant difference is observed by taking a smaller time-step. The statistics of sequences of visible jumps and their intertransition times were obtained by analyzing their frequency from long enough trajectories such that the empirically measured quantities presented reasonable convergence.
The parameter $k$, which sets the time-scale of the transition rates in Eq. \eqref{EqRates1} and Eq. \eqref{EqRates2}, 
and the period $T$ are set to $k=10$ and $T=1$. We here consider that only the transitions between states $1$ and $2$ are 
visible, i.e., $\mathcal{L}=\{1\to2,2\to1\}$.

\subsection{Energy, energy barrier and  entropy production estimate}

\begin{figure}
\subfigure[]{\includegraphics[width=8cm]{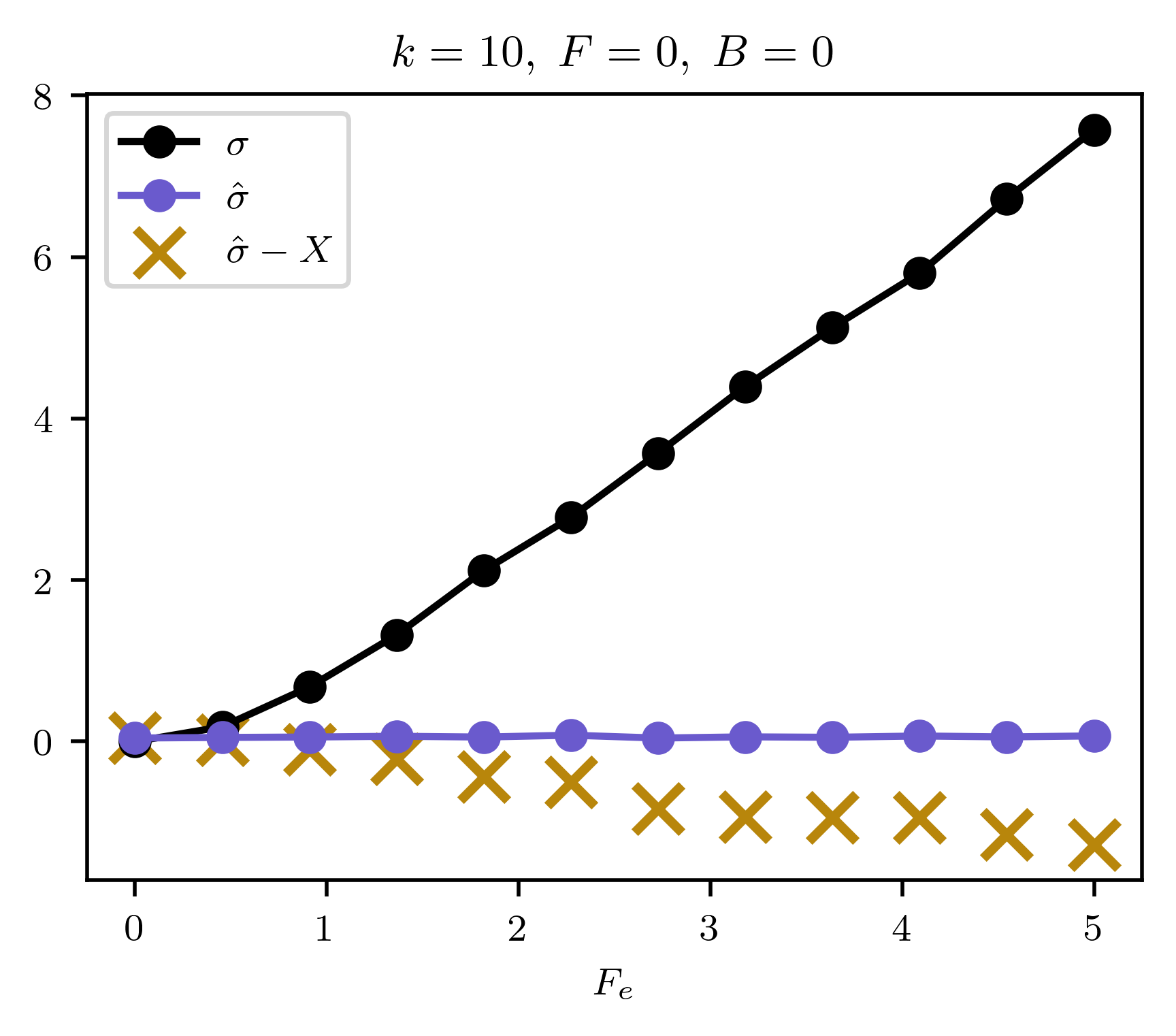}\label{fig3a}}
\subfigure[]{\includegraphics[width=8cm]{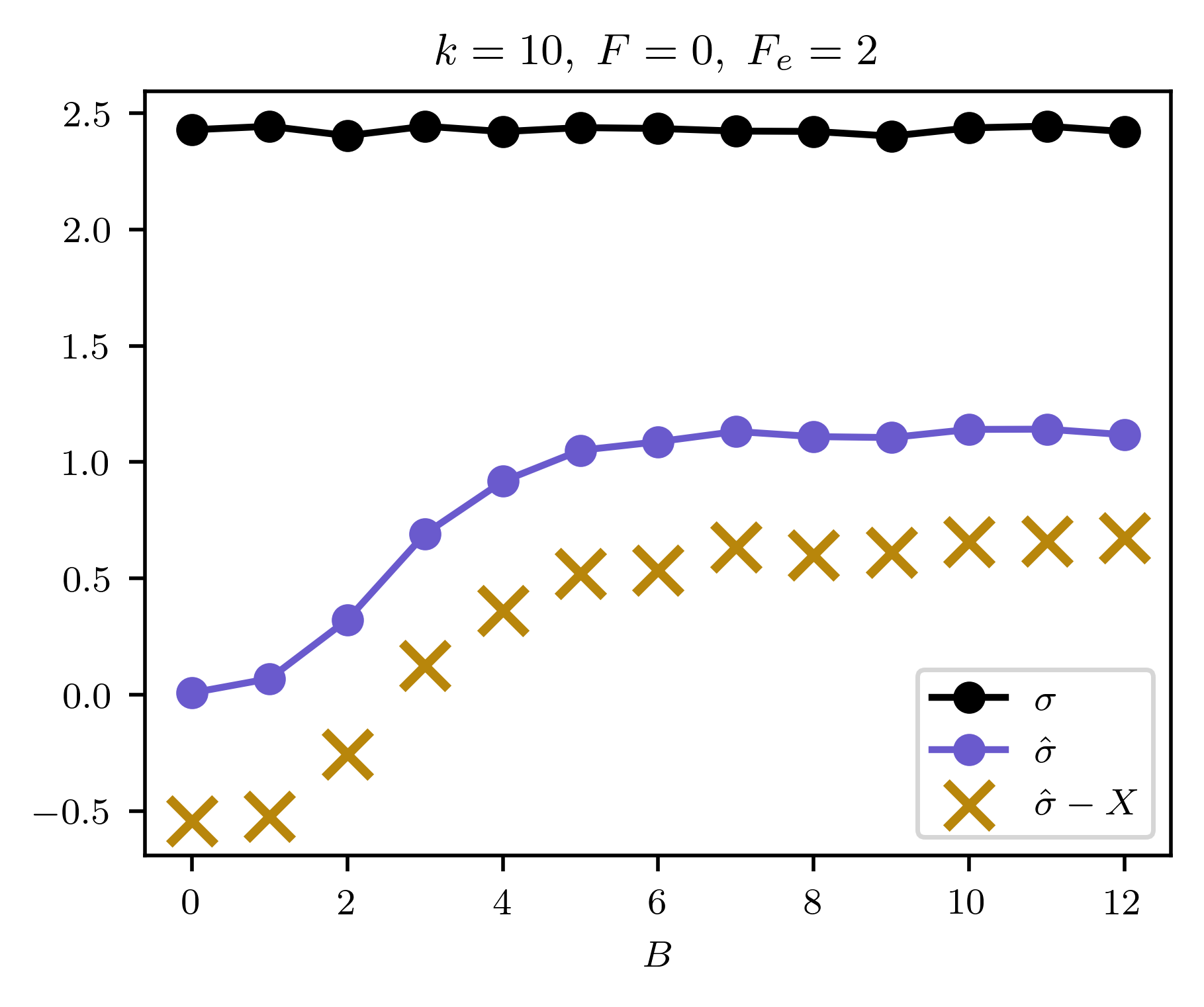}\label{fig3b}}
\vspace{-2mm}
\caption{Entropy production $\sigma$, its estimate $\hat{\sigma}$ and its lower bound $\hat{\sigma} - X$ as functions of (a) the energy $F_e$ for $B=0$ and (b) the energy barrier $B$ for $F_e=2$.}   
\label{fig3} 
\end{figure}

In Fig. \ref{fig3}, we show results for the fixed external affinity $F=0$. In Fig. \ref{fig3a} the energy barrier is $B=0$, which leads to a net current $J=0$, as
discussed in Sec. \ref{sec2}.  Even though the current $J$ does not appear in the formula for $\sigma$ in 
Eq. \eqref{EqEnt1} it seems to play a fundamental role for the estimate $\hat{\sigma}$. For $B=0$, we see that the estimate $\hat{\sigma}$ is numerically 
compatible with $0$ and, therefore, does not provide any useful information about the rate of entropy production $\sigma$. 
Hence, the emergence of a current $J$, which is not necessary for a non-zero entropy production $\sigma$ in periodically driven systems, 
seems to be a necessary condition for a meaningful estimate $\hat{\sigma}$.  

The case of a non-zero energy barrier is shown in Fig \ref{fig3b}. Here we see that the estimate $\hat{\sigma}$ is non-zero for $B>0$ that leads to the emergence 
of a non-zero current $J$. From the results in Fig. \ref{fig3}, we observe that even for unicyclic networks the estimate $\hat{\sigma}$ is not equal to the 
entropy production $\sigma$. This situation is in contrast to steady states, where a similar estimate becomes equal to the entropy production for unicyclic networks \cite{meer22,haru22}. 

In summary, our results for the case of $F=0$ lead to two main conclusions. First, it seems that the emergence of a current $J$ is a relevant condition for the usefulness of $\hat{\sigma}$ as an
estimator. Second, even in a unicyclic network the estimate $\hat{\sigma}$ is not equal to the exact rate of entropy production $\sigma$.

\subsection{Role of external fixed affinity $F$}

\begin{figure}
\subfigure[]{\includegraphics[width=8cm]{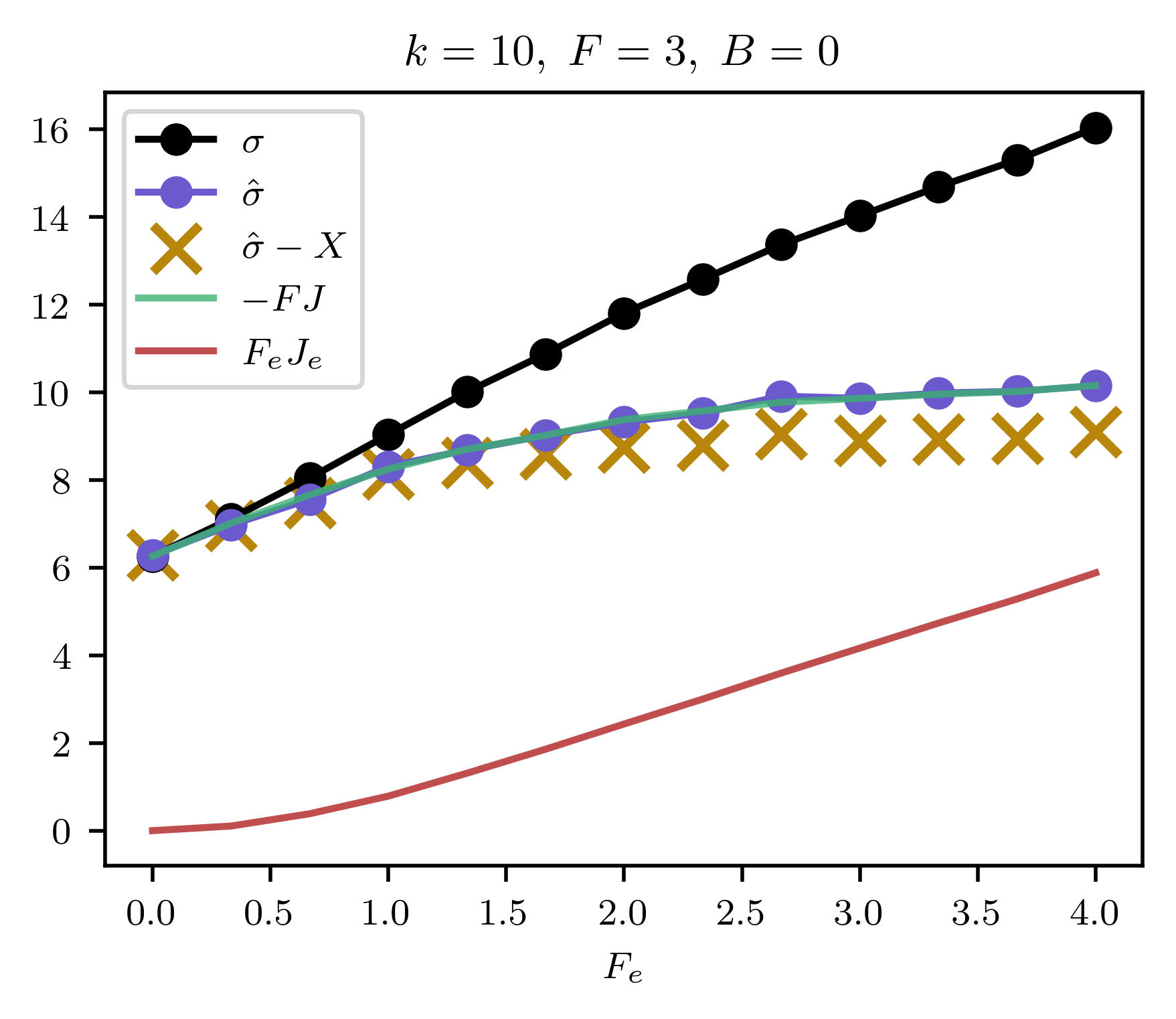}\label{fig4a}}
\subfigure[]{\includegraphics[width=8cm]{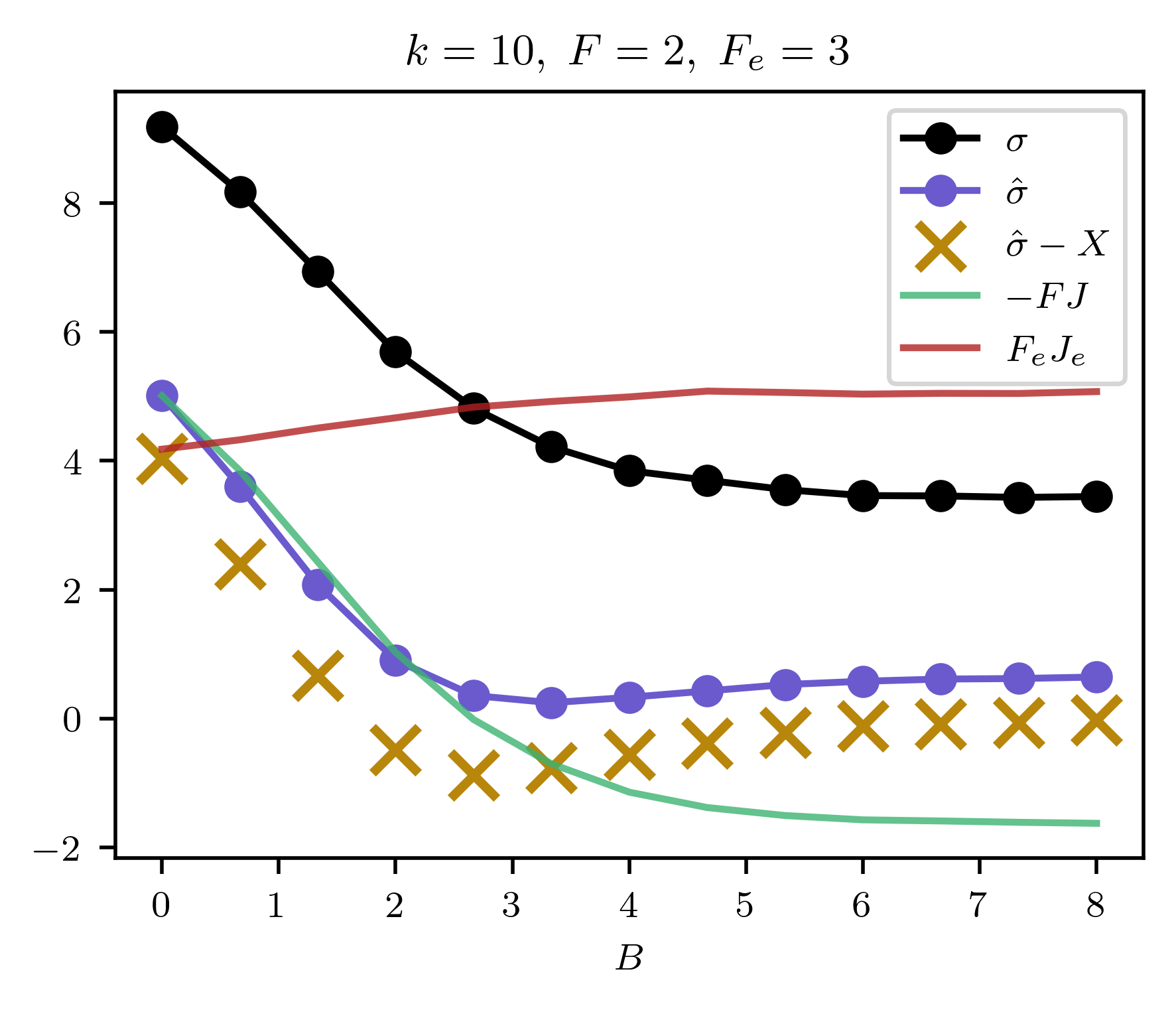}\label{fig4b}}
\vspace{-2mm}
\caption{Entropy production $\sigma= FJ+F_eJ_e$, its estimate $\hat{\sigma}$ and its lower bound $\hat{\sigma} - X$ as functions of (a) the energy $F_e$ for $B=0$ and (b) the energy barrier $B$ for $F_e=3$.} 
\label{fig4} 
\end{figure}

We now consider the case of a non-zero fixed affinity $F$. 
In Fig.~\ref{fig4a} we compare the estimate $\hat{\sigma}$ for the energy barrier $B=0$ with the rate of entropy production $\sigma$, 
and its two contributions in Eq.~\eqref{EqEnt1}. As we can see in the figure, $\hat{\sigma}$
 seems to follow the term $-FJ$ and does not capture much information about the contribution due to the work term $J_eF_e$, reinforcing the connection between the estimator and the net current.

In Fig.~\ref{fig4b} we show our results for $B\neq0$. Here, we can see that the estimate $\hat{\sigma}$ is sensitive to the work term $J_eF_e$, even 
in the regime where $-JF$ becomes negative. In this regime, the systems operates as an engine with work exerted against the internal force $F$.
Therefore, the results in Fig.~\ref{fig4}, together with the need for net motion, suggest that the presence of energy barriers seems to be a necessary condition for the estimate $\hat{\sigma}$
to capture information on the work term $F_eJ_e$ of the rate of entropy production $\sigma$.

\section{Proof of the bound and the emergence of the extra term}
\label{sec4}

\subsection{First  bound $\sigma_\mathcal{L}$}

A full trajectory of the Markov process with all transitions visible is denoted by $\gamma_{t_f}$, where $t_f$ is the total time of the trajectory. 
We are interested in the limit $t_f\to \infty$. The rate of entropy production is given by the relative entropy 
between the probability of a trajectory \(P[\gamma_{t_f}]\) and its time reverse under reversed protocol 
\(P^\dagger[\gamma_{t_f}^\dagger]\) \cite{seif12}, where the superscript in $P^\dagger$
means the probability associated with the reversed protocol and the superscript in  $\gamma_{t_f}^\dagger$ means the reverse of the trajectory $\gamma_{t_f}$. 
This formula is written as 
\begin{equation}
\sigma = \lim_{{t_f}\to\infty} \frac{1}{{t_f}} \sum_{\gamma_{t_f}} P[\gamma_{t_f}] \ln \frac{P[\gamma_{t_f}]}{P^\dagger[\gamma_{t_f}^\dagger]},
\end{equation}
Formally, the sum over all trajectories $\sum_{\gamma_{t_f}}$ corresponds to a functional integration.

Let \(\theta\) be a coarse-graining map such that \(\Gamma_{t_f} = \theta \gamma_{t_f}\), it can represent arbitrary coarse-graining. 
Here we consider the removal of hidden transitions and only a subset of transitions $\mathcal{L}$ remain visible. 
This map is non-injective (many-to-one) since the trajectory details are contracted, thus many full trajectories  $\gamma_{t_f}$ 
can give rise to the same coarse-grained trajectory $\Gamma_{t_f}$. For the probability of the coarse-grained trajectory 
$\Gamma$ we can write
\begin{equation}
P[\Gamma] = \sum_{\gamma \in \theta^{-1}\Gamma} P[\gamma].
\end{equation}
We consider a time-reversal operation that commutes with the coarse-graining map, \((\theta \gamma)^\dagger = \theta( \gamma^\dagger ) \eqqcolon \Gamma^\dagger\), which is achieved by noticing that transitions $\ell$ are replaced by $\bar{\ell}$ in the time-reversed trajectory. A lower bound on entropy production rate $\sigma$, accessible from the statistics of the visible transition that pertain to $\mathcal{L}$, is defined as
\begin{equation}
\label{vEPR}
\sigma_\mathcal{L} \equiv \lim_{{t_f}\to\infty} \frac{1}{{t_f}} \sum_{\Gamma_{t_f}} P[\Gamma_{t_f}] \ln \frac{P[\Gamma_{t_f}]}{P^\dagger[\Gamma_{t_f}^\dagger]},
\end{equation}
which fulfills
\begin{equation}
\sigma \geq \sigma_\mathcal{L},
\end{equation}
due to the log-sum inequality.
 
The coarse-grained trajectory $\Gamma_{t_f}$ is a sequence of visible transitions $\ell_i\in \mathcal{L}$ that take place at times $t_i$. If the trajectory has $N$ visible transitions then $i=1,2,\ldots,N$. The probability of a given trajectory \(\Gamma\) reads
\begin{equation}
\label{traj_prob}
P[\Gamma] = \PP_{\ell_1,t_1} \prod_{i=2}^N \PP_{\ell_i, t_i \vert \ell_{i-1},t_{i-1}},
\end{equation}
where $\PP_{\ell_i, t_i \vert \ell_{i-1},t_{i-1}}$ is the conditional probability of transition \(\ell_i\) being observed at \(t_i\) given that transition \(\ell_{i-1}\) occurred at \(t_{i-1}\), provided no other visible transitions in $\mathcal{L}$ occurred in between. A similar expression is valid for $P^{\dagger}[\Gamma^{\dagger}]$, which is
\begin{equation}
\label{traj_probrev}
P^{\dagger}[\Gamma^{\dagger}] = \PP^\dagger_{\bar{\ell}_N,t_f-t_N} \prod_{i=0}^{N-2} \PP^\dagger_{ \overline{\ell}_{N-i-1},t_f-t_{N-i-1} \vert \overline{\ell}_{N-i},t_f-t_{N-i}}.
\end{equation}
Even though we are interested only in the limit of $t_f\to\infty$, these two expressions for the probability of a trajectory are valid for any finite $t_f$.

We now obtain an expression for the lower bound on the entropy production in Eq. \eqref{vEPR} using the expressions for the probability of a 
trajectory in Eq.~\eqref{traj_prob} and in Eq.~\eqref{traj_probrev}, which reads
\begin{equation}
\sigma_\mathcal{L} =  \lim_{{t_f}\to\infty} \frac{K}{t_f} \sum_{\ell', \ell\in\mathcal{L}} \int_{t}^{t_f} \mathrm{d}t' \int_0^{t_f} \mathrm{d}t
\PP_{\ell',t';\ell,t} \ln \frac{\PP_{ \ell',t'\vert \ell,t}}{\PP^\dagger_{\overline{\ell},{t_f}-t\vert  \overline{\ell'},{t_f}-t'}},
\label{EqsigmaK2}
\end{equation}
where $\PP_{\ell',t';\ell,t}= \PP_{\ell',t'\vert \ell,t} \PP_{\ell,t}$ is the joint distribution. 
In this expression, since we are considering the limit  $t_f\to\infty$, the contribution of the boundary term in $\ln (P[\Gamma_{t_f}]/P^\dagger[\Gamma_{t_f}^\dagger])$, 
which is $\ln(\PP_{\ell_1, t_1}/\PP^\dagger_{\bar{\ell}_N, t_f-t_N} )$, goes to zero. 

\begin{figure}
    \includegraphics[width=8cm]{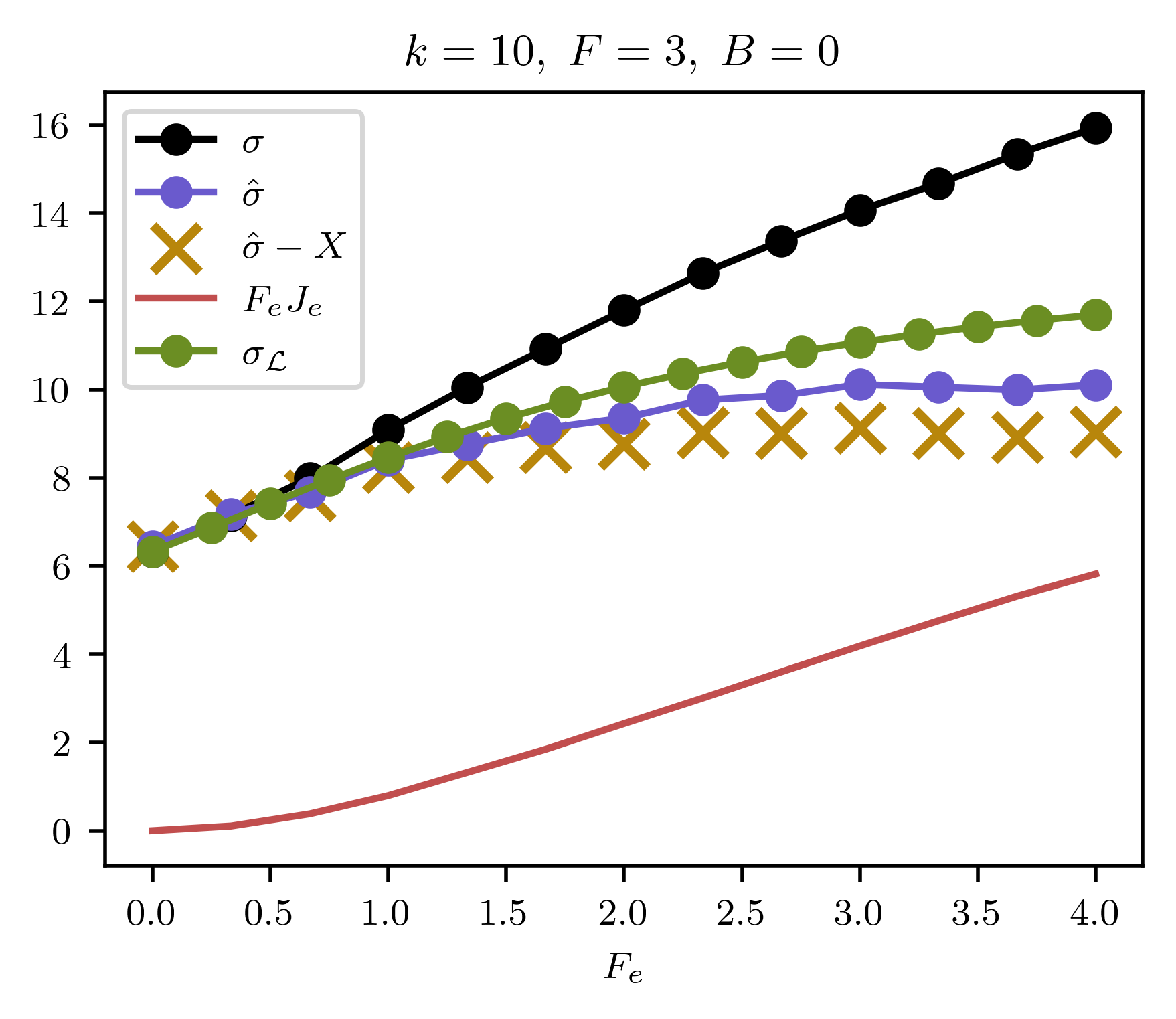}
\caption{The quantities of Fig.~\ref{fig4a} and $\sigma_\mathcal{L}$.} 
\label{fig_sigmaL} 
\end{figure}

The expression in \eqref{EqsigmaK2} can be further simplified for a time-periodic system with period $T$, this simplification reads
\begin{equation}
\label{EqsigmaK3}
\sigma_\mathcal{L} =   \frac{K}{T} \sum_{\ell', \ell\in\mathcal{L}} \int_{0}^{\infty} \mathrm{d}\tau \int_0^{T} \mathrm{d}t_0 \PP_{\ell',{\tau+t_0}; \ell,t_0} \ln \frac{\PP_{\ell',{\tau+t_0}\vert \ell,t_0}}{\PP^\dagger_{\overline{\ell}, \tau+t_0^*  \vert \overline{\ell'},t_0^*}},
\end{equation}
where $t_0^*= T-[(\tau+t_0) \mod T]$. In Fig.~\ref{fig_sigmaL} we show that  $\sigma_\mathcal{L}$  can provide a better estimate of $\sigma$ in comparison to $\hat{\sigma}$, where $\sigma_\mathcal{L}$ was calculated using the analytical methods of Section~\ref{sec5}. However, in practice, it is much more difficult to obtain this estimator from a trajectory generated in a simulation or experiment. The main difference is that, for each pair of transitions, this quantity requires one histogram for each initial time $t_0$ in the period, while $\hat{\sigma}$ uses only one histogram.

\subsection{Elimination of $t_0$}

Our estimate $\hat{\sigma}$ that only depends on the intertransition times arises with the use of a log-sum inequality in $t_0$.  The extra term is defined as
\begin{equation}
\label{EqDefX}
X\equiv\frac{K}{T} \sum_{\ell', \ell\in\mathcal{L}} \int_{0}^{\infty} \mathrm{d}\tau \int_0^{T} \mathrm{d}t_0 P_{\ell',{\tau+t_0}; \ell,t_0} \ln 
\frac{P_{t_0\vert \ell}}{P^\dagger_{t_0^*\vert \overline{\ell'}}} \geq 0.
\end{equation}
From this definition of $X$ and \eqref{EqsigmaK3}, using the log-sum inequality in $t_0$ we obtain 
\begin{equation}
\sigma_\mathcal{L}+X\ge \hat{\sigma}, 
\end{equation}
where $\hat{\sigma}$ is our estimate given by Eq. \eqref{vEPR_FPT}. Since 
$\sigma\ge \sigma_\mathcal{L}$ we obtain
\begin{equation}
\sigma\ge \hat{\sigma}- X.
\end{equation}
 The extra term $X$ fulfills the inequality 
\begin{equation}
 X\ge 0,
\end{equation}
which is obtained from Eqs. \eqref{EqPtgivenl}, \eqref{EqDefX}, and the log-sum inequality in $t_0$. The term  $X$ can be calculated from 
a trajectory (and another one with the reversed protocol) in the following way. 
First, we obtain the distributions $\PP_{t_0\vert \ell}$ from a 
trajectory with forward protocol and $\PP^\dagger_{t_0\vert \ell}$ from a trajectory with reversed protocol. 
Then we can run over all transitions in a 
visible trajectory $\Gamma$ with forward protocol, with its probability represented in Eq. \eqref{traj_prob}, 
and compute $\ln ( \PP_{t_i\vert \ell_i} / P^\dagger_{t_f-t_i\vert \overline{\ell}_i} )$ whenever a visible transition $\ell_i$ happens at time $t_i$. If the probabilities are empirically
inferred from the same trajectory, correlations can give rise to convergence issues, thus it
is better to use independent trajectories.
Note that there is no need to calculate a distribution that depends on two times in order to calculate $X$ from a trajectory, which was a main practical issue with the tighter 
estimate $\sigma_\mathcal{L}$.

In summary, our estimate $\hat{\sigma}$ is not connected to $\sigma$ by a formal inequality to our knowledge. However, within our numerics restricted to 
one case study $\hat{\sigma}$ is below $\sigma$. This evidence is not enough to make a conjecture but it leaves the possibility of an inequality open.
 If we consider $\hat{\sigma}-X$ as an estimate, then we do have a formal inequality.

\section{Analytical method to determine the intertransition time distribution}
\label{sec5}

Our estimator $\hat{\sigma}$ depends on the intertransition time probability density $\psi_{\ell'\vert\ell}(\tau)$, which can be obtained from histograms of observable trajectories. 
We show how to calculate this distribution analytically for a periodically driven system. Similar to a procedure for steady states from \cite{meer22,haru22, sekimotoFPT}, 
we map the problem of determining $\psi_{\ell'\vert\ell}(\tau)$ onto a survival probability problem of an auxiliary Markov process with absorbing states. The difference here is that 
the transition rates are time-dependent, which makes the procedure more involved.

The formal solution of the master equation \eqref{EqMaster} is
\begin{equation}\label{eq:me_evolution}
	P_i(t) = \sum_j \left[ \mathcal{T} \left\lbrace \exp \int_{t_0}^t \mathrm{d}s \mathbf{W} (s) \right\rbrace \right]_{j,i} P^{\textrm{ini}}_j (t_0),
\end{equation}
where $t_0$ is the initial time, $P^{\textrm{ini}}$ the initial distribution and $\mathcal{T} \lbrace \exp \bullet \rbrace$ denotes the time-ordered matrix exponential. The stochastic matrix $\mathbf{W}(t)$ has elements   $[\mathbf{W} (s)]_{ij}=w_{ji}(s)$ for $i\neq j$ and $[\mathbf{W} (s)]_{ii}=-\sum_j \omega_{ij}(s)$. The long time solution $P_i(t)$ of the 
master equation can be obtained with Floquet theory \cite{bara18}.

To calculate the intertransition time distribution $\psi_{\ell'\vert\ell}(t)$ we define an auxiliary process that has absorbing states. This auxiliary process has an extra number of states that equals the number of transitions in $  \mathcal{L}$, the set of visible transitions. All extra states are absorbing. For instance, if the transition $\ell=1\to2\in \mathcal{L}$  then the transition from $1$ to $2$ does not go to state $2$ in this new auxiliary process, it goes to a new absorbing state denoted $\ell=1\to2$. The 
stochastic matrix associated with this auxiliary process is denoted $\mathbf{W}_\text{aux}(t)$. 

As an example, we consider the three-state model. The stochastic matrix associated with the original process reads
\begin{gather}
\mathbf{W}(t)= \begin{bmatrix}
   -r_1(t) &   w_{21}(t) &  w_{31}(t)\\
    w_{12}(t)&   -r_2(t) &  w_{32}(t)\\
      w_{13}(t) &   w_{23}(t) &  -r_3(t)
\end{bmatrix},
\end{gather}
where $r_i(t)= \sum_j w_{ij}(t)$.  For $\mathcal{L}=\{1\to2,2\to1\}$, the stochastic matrix for the auxiliary process reads
 \begin{gather}
\mathbf{W}_\text{aux}(t)=
\begin{bmatrix}
   -r_1(t)      &   0             &  w_{31}(t) & 0              & 0\\
    0             & -r_2(t)       &  w_{32}(t) & 0 & 0\\
   w_{13}(t) &   w_{23}(t) &  -r_3(t)      & 0             & 0\\
           0      &   w_{21}(t)             &  0              & 0             & 0\\
           w_{12}(t)      &   0             &  0              & 0             & 0
\end{bmatrix},
\end{gather}
where the fourth state is associated with $2\to1$ and the fifth state is associated with $1\to2$.

\begin{figure}
\includegraphics[width=\columnwidth]{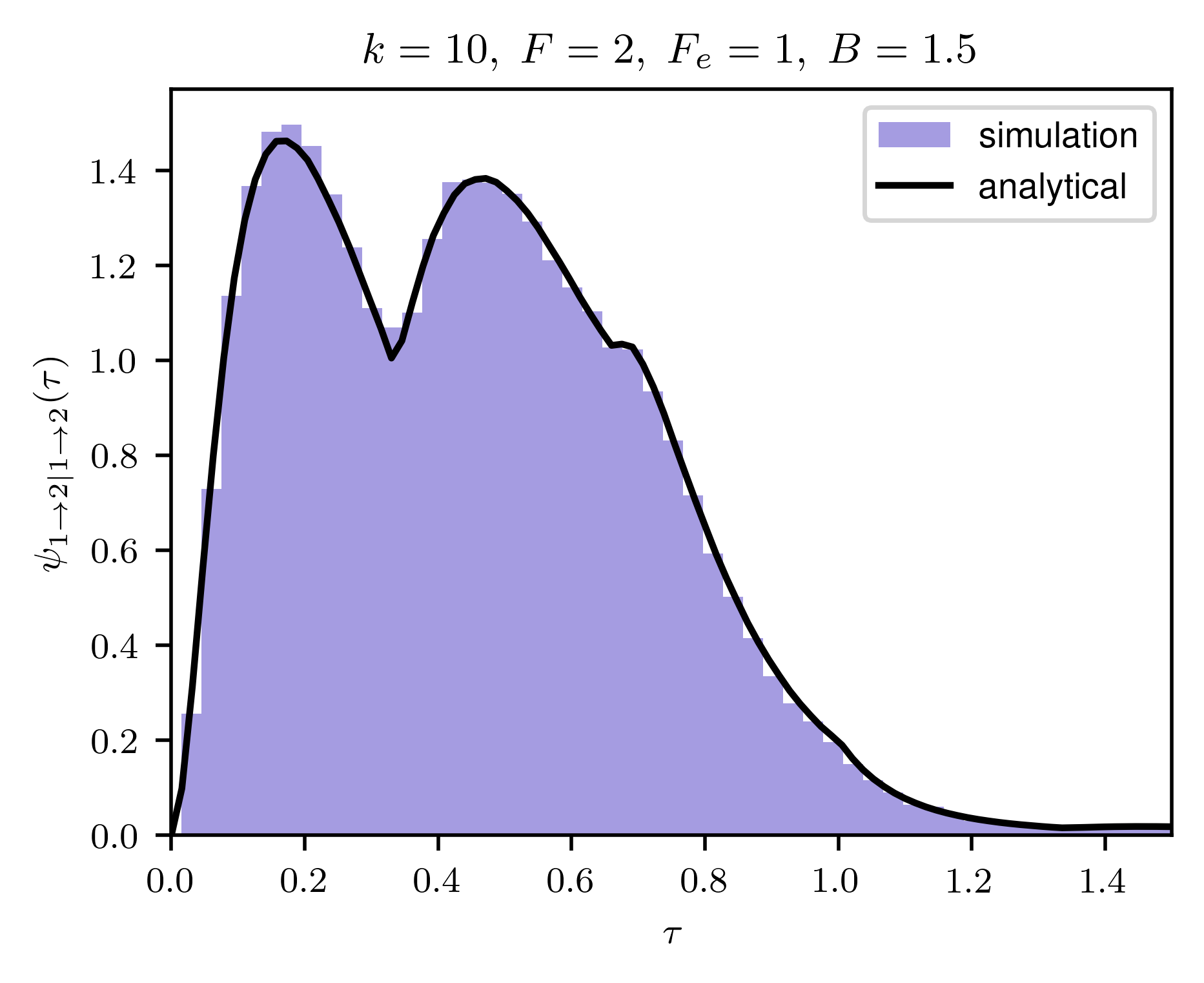}
\vspace{-2mm}
\caption{Comparison between analytical method and numerics for the intertransition time distribution $\psi_{\ell'|\ell}(\tau)$ for the three-state model. 
The transitions are  $\ell=\ell'=12$. }   
\label{fig5} 
\end{figure}

The conditional distribution $\PP_{\ell',\tau+t_0\vert\ell,t_0}$ is related to the time-dependent solution of this auxiliary process. Let's denote a generic state of the auxiliary process by $a$. The probability of being in state $a$ at time $t$ is denoted by $Q_a(t)$. The condition that $\ell$ and $t_0$ are given in the conditional probability $\PP_{\ell',\tau+t_0\vert\ell,t_0}$ is reflected in the initial probability of this auxiliary process. The time for the transition rates for this initial condition is $t_0$. If the transition $\ell=i\to j$, 
then the initial state of the auxiliary process is $j$, i.e., the initial probability density of the auxiliary process $Q^{\textrm{ini}}_a (t_0)$ is a delta function that is one only for the state corresponding to $j$ and zero otherwise. Hence, we obtain  $\PP_{\ell',\tau+t_0\vert\ell,t_0}= \partial_\tau Q_a(\tau+t_0)$, where the state $a$ is the absorbing state corresponding to the transition $\ell'$ and 
\begin{equation}\label{eq:me_evolutionaux}
Q_a(\tau+t_0) = \sum_b \left[ \mathcal{T} \left\lbrace \exp \int_{t_0}^{\tau+t_0} \mathrm{d}s \mathbf{W}_\text{aux} (s) \right\rbrace \right]_{b,a} Q^{\textrm{ini}}_b (t_0).
\end{equation}

The intertransition time distribution  $\psi_{\ell'|\ell}(\tau)$ can be calculated by evaluating $Q_a(\tau+t_0)$ from the above equation for all $t_0\in [0,T]$, taking its derivative and then applying equation \eqref{eq:psi}. In Fig. \ref{fig5}, we show the agreement between the intertransition time distribution obtained analytically from this method and numerically from a single trajectory.

\section{Conclusion}
\label{sec6}

We proposed a method for the inference of entropy production from the statistics of the intertransition times 
of a few visible transitions in periodically driven systems. If sufficient data is available, the estimator $\sigma_\mathcal{L}$, which depends on both the intertransition time and 
the initial time of the first transition estimator, provides the tightest bound to the rate of entropy production. The more accessible estimator $\hat{\sigma}$ is obtained by eliminating the need to track the initial time, it is related to the rate of entropy production $\sigma$ with a formal inequality that requires the extra term $X$. While our numerics indicate the possibility of the inequality $\sigma\ge\hat{\sigma}$, this result remains restricted to our case study. The formal inequality we obtained here is $\sigma\ge \hat{\sigma}-X$.
The emergence of this extra term and 
the necessity to generate two trajectories, one with forward protocol and another with backward protocol, are two main differences between 
the inference of entropy production for steady states and periodically driven systems.
 
We applied our estimators to a simple three-state model for a molecular pump leading to two general lessons. First, even in a unicyclic 
system the estimate does not equal the exact rate of entropy production. This result shows a key difference in relation to steady states, for which
the estimate does equal the exact entropy production in unicyclic systems \cite{meer22,haru22}. Second, the emergence of a net current in 
periodically driven systems, which is not a necessary condition for a non-zero entropy production, seems to be a necessary condition 
for $\hat{\sigma}$ to provide a meaningful estimate of the rate of entropy production. We observed that when the net current is zero the estimate 
is often numerically compatible with zero and does not capture any information about the real rate of entropy production. It would be interesting to investigate in which regimes the estimator $\hat{\sigma}$ is exactly zero, for instance in the case of no-pumping. For cases with an emergent current, 
such as a molecular pump with a non-zero energy barrier or for a non-zero thermodynamic affinity,  the method can provide a good estimate of the rate 
of entropy production. Hence, the method we propose here is a good candidate to estimate the rate of entropy production in physical systems with a 
non-zero net current such as molecular pumps and should not give much information for systems with a zero net current, such as heat engines.

A recent independent study \cite{maier24} on the same topic of inference of entropy production in periodically driven systems came to our knowledge after we completed this work. Their approach is complementary to ours: They propose an estimator based on intertransition time statistics similar to $\hat{\sigma}$, in the form of a numerical conjecture, that has the advantage of only depending on statistics associated with the forward trajectory. Our work shows that, at the expense of running trajectories with the reverse protocol, a formal inequality between estimator, extra term $X$ and real rate of entropy production can be proven. Moreover, it would be interesting to compare the performance of our present bounds with the bound for periodically driven systems associated with the thermodynamic uncertainty relation from \cite{koyu20}, which depends on a single current. Such a comparison can be meaningful if the visible transitions are the ones related to a single current.
 
From a mathematical perspective, it would be interesting to prove whether the inequality $\sigma\ge \hat{\sigma}$, observed to hold within our numerics, is indeed correct. Interesting directions for future work include the study of inference of entropy production for the case of cyclic stochastic protocols, the investigation of other methods that do not require access to a trajectory with backward protocol, and a rigorous classification of physical systems for which the method can provide a good estimate.

\section{Acknowledgments}

PH was supported by the project INTER/FNRS/20/15074473 funded by F.R.S.-FNRS (Belgium) and FNR (Luxembourg). CEF acknowledges the financial support from brazilian agencies CNPq and FAPESP under grants 2024/03763-0 and 2022/15453-0.

\bibliographystyle{apsrev4-1}


%


\end{document}